\documentclass[prl,twocolumn,showpacs,superscriptaddress]{revtex4}
\usepackage{graphicx,amssymb,amsmath,color,bm}
\begin{document}

\title{Dynamics of the (spin-) Hall effect in topological insulators and graphene}
\author{Bal\'azs D\'ora}
\email{dora@kapica.phy.bme.hu}
\affiliation{Department of Physics, Budapest University of Technology and
  Economics, Budafoki \'ut 8, 1111 Budapest, Hungary}
\author{Roderich Moessner}
\affiliation{Max-Planck-Institut f\"ur Physik komplexer Systeme, N\"othnitzer Str. 38, 01187 Dresden, Germany}
\date{\today}

\begin{abstract}
A single two-dimensional Dirac cone with a mass gap produces a quantized (spin-) Hall step in the absence of magnetic field.
What happens in strong electric fields? This question is investigated by analyzing time evolution and dynamics of the (spin-) Hall effect.
After switching on a longitudinal electric field, a stationary Hall current is reached through damped oscillations. 
The Hall conductivity remains quantized as long as the electric field ($E$) is too weak to induce Landau-Zener transitions, but
quantization breaks down for strong fields and  the conductivity decreases as $1/\sqrt{E}$.
These apply to the (spin-) Hall conductivity of graphene and the Hall and magnetoelectric response of topological insulators.
\end{abstract}


\maketitle

The unique electronic
properties of graphene can be traced back to the the pseudo-relativistic Dirac equation and its linear energy dispersion with zero bandgap. 
It exhibits a plethora of interesting and fascinating physical phenomena related to electric and heat transport, magnetic field effects, valley and spintronics\cite{castro07}.
The `half-integer' quantum Hall effect, in spite of its environmental fragility, has been observed at room temperature\cite{roomhallgraphene} due to the unusual Landau quantization of Dirac electrons.
With spin-orbit coupling taken into account, graphene in principle realizes a spin-Hall
insulator\cite{kanemele1}, belonging to the class of topological insulators
(TI). 

Moreover, Dirac electrons also occur as surface states of three-dimensional
TI\cite{hasankane,bernevig,phystoday}, close relatives of integer quantum Hall
states. These materials are predicted to display a variety of peculiar phenomena,
such as spin- and surface quantum Hall effects and the closely related
topological magnetoelectric effect\cite{tme}, allowing for the control of
magnetization by electric field. As opposed to the even number of Dirac cones
in graphene, three dimensional TI can have an odd number of Dirac cones on a
surface.  
Due to time reversal symmetry, these states are robust with respect to non-magnetic disorder, similarly to how pair breaking in s-wave superconductors is prohibited by
potential scatterers.

The hallmark of (pseudo-) relativistic massive Dirac electrons\cite{ludwig94} is a single quantum Hall step around half-filling in the absence of magnetic field between $\sigma_{xy}=\pm e^2/2h$.
We ask how this picture gets modified in the presence of strong electric fields.
Generally, a driving electric field can produce a sizable density of electron and hole excitations around the Dirac point in a highly 
non-thermal, non-stationary momentum distribution\cite{doraschwinger}.
Consequently, the longitudinal transport of Dirac electrons features
Klein-tunneling\cite{beenakker} and Schwinger's pair
production\cite{schwinger,allor,doraschwinger,rosenstein} in a stationary or
time-dependent framework, 
when the electric field is represented by a static scalar or a time-dependent vector potential, respectively.
The latter approach directly yields the non-equilibrium momentum distribution and the time-dependent current at finite electric fields. 
While it does not use any kind of equilibrium or out of
equilibrium response formalism (Kubo/Landauer), it still reproduces known
results and makes predictions for the non-linear behaviour of the electric current as an 
example\cite{doraschwinger,mauri}.
On the other hand, a strong electric field alters not only the longitudinal transport\cite{doraschwinger}, but is expected to
modify the transverse conductivity, involving the non-equilibrium quantum (spin-) Hall breakdown.
Common wisdom tells that while there are no power law
corrections to the integer Hall conductivity for weak electric fields, with its
quantization `topologically' protected, there can be exponentially small
corrections. When these grow with field, quantization breaks down.

To consider the problem in detail, we elaborate on the time evolution of the Hall current for massive Dirac Fermions, after switching on a \textit{longitudinal} electric field.
 We show that a {\em stationary transverse}  current develops for long times, characterized by a quantized Hall conductivity for weak fields, 
crossing over to a strongly field dependent Hall response with increasing field.
This result applies to the quantum (spin-) Hall breakdown of
graphene\cite{singh} as well as for the related\cite{hasankane} surface Hall
and magnetoelectric effect in TI.

The low energy description around the $K$ point in the Brillouin zone of
graphene\cite{castro07} or on the surface state of  a 3D
TI\cite{franz,hasankane} (after a $\pi/2$ rotation of the spin around $\hat z$),
 in the presence of a uniform, constant electric field ($E>0$) in the $x$ direction 
[switched on at $t=0$, through a time dependent vector potential ${\bf A}(t)=(A(t),0,0)$ with $A(t)=Et\Theta(t)$] is written as 
\begin{gather}
i\hbar\partial_t\Psi_p(t)=\left(v_F(p_x-eA(t)),v_Fp_y,\Delta\right)\cdot{\bm\sigma}\Psi_p(t),
\label{diraceq}
\end{gather}
where $v_F$ is the Fermi velocity, and the Pauli matrices ($\sigma$) encode the two sublattices\cite{castro07} of the honeycomb 
lattice in graphene, or the physical spin in TI. $\Delta>0$ is the mass gap,
originating from the intrinsic spin-orbit coupling (SOC) in
graphene\cite{kanemele1}, or from a thin ferromagnetic film covering the
surface of TI, lifting the Kramer's degeneracy of the Dirac point.

To make our analysis more transparent, we perform a two-step unitary transformation, $U=U_1U_2$. Firstly, a time independent rotation around the $\sigma_x$ axis as
$U_1=C_+-i\sigma_xC_-$,
with
$C_\pm=[1\pm (v_Fp_y/\sqrt{(v_Fp_y)^2+\Delta^2)}]^{1/2}/\sqrt 2$,
and secondly a time dependent one, bringing us to the adiabatic basis
as
$U_2=\exp(-i\varphi(t)\sigma_z/2)[\sigma_x+\sigma_z]$
with $\tan \varphi(t)=\sqrt{p_y^2+(\Delta/v_F)^2}/[p_x-eA(t)]$.
The resulting instantaneous energy spectrum in the upper Dirac cone is 
$\varepsilon_p(t)=\sqrt{\Delta^2+v_F^2((p_x-eA(t))^2+p_y^2)}$. 
The transformed time dependent Dirac equation reads as
\begin{gather}
i\hbar\partial_t\Phi_p(t)=\left[\sigma_z\varepsilon_p(t)-\sigma_x\frac{\hbar 
v_FeE\sqrt{(v_Fp_y)^2+\Delta^2}}{2\varepsilon^2_p(t)}\right]\Phi_p(t),
\label{diractrans}
\end{gather}
and $\Psi_p(t)=U\Phi_p(t)$, with initial (ground-state) condition $\Phi_p^T(t=0)=(0,1)$,  in which the lower (upper) Dirac cone is fully occupied (empty).
The electric field alters  the energy spectrum and induces off-diagonal terms in the Hamiltonian.
Two energy scales at the moving Dirac point (${\bf p}=(eEt,0)$) in Eq. \eqref{diractrans}, characterise the low-energy physics: the diagonal energy ($\Delta$) and off-diagonal coupling ($\hbar v_FeE/2\Delta$), which triggers transitions between the two gap edges or levels using Landau-Zener terminology\cite{vitanov}. A crossover from weak to strong field is thus expected
at $E\sim \Delta^2/\hbar v_Fe$, irrespective of the explicit value of $t$, as we confirm below by a more detailed analysis.

The quantity we focus on is the time dependent transverse charge current, $j_y=-ev_F\sigma_y$ in the basis of Eq. \eqref{diraceq}, with spin current and conductivity differing only by a factor $\hbar/e v_F$. 
For TI, $j_y$ coincides with the topological magnetic field induced \textit{parallel} to the applied electric field 
(after the $\pi/2$ rotation of the spin, leading to Eq. \eqref{diraceq}) and monitors the magnetoelectric effect\cite{tme,franz}.

By denoting 
$\Phi_p^T(t)=(\alpha_{p}(t),\beta_{p}(t))$, 
charge conservation implies $|\beta_{p}(t)|^2=1-n(t)$, where $n_{p}(t)=|\alpha_p(t)|^2$ is the number of electrons which have  tunneled into the initially empty
upper Dirac cone. After  multiplying the transformed Dirac equation, Eq.  \eqref{diractrans}
with $\Phi_p^+(t)\sigma_x$ or $\Phi_p^+(t)$ from the left, we get
\begin{gather}
\langle j_y\rangle_p(t)=-\frac{ev_F\Delta\hbar}{\varepsilon_p(t)}\left(\frac{\partial_t\left[\varepsilon^2_p(t)\partial_tn_p(t)\right]}{v_FeE((v_Fp_y)^2+\Delta^2)}+\right.\nonumber\\
\left.+\frac{v_FeE}{2\varepsilon^2_p(t)}\left(2 n_p(t)-1\right)\right),
\label{expr1}
\end{gather}
which depends only on $n_p(t)$ and its time derivatives.
\begin{figure}[h!]
{\includegraphics[width=5cm,height=5cm]{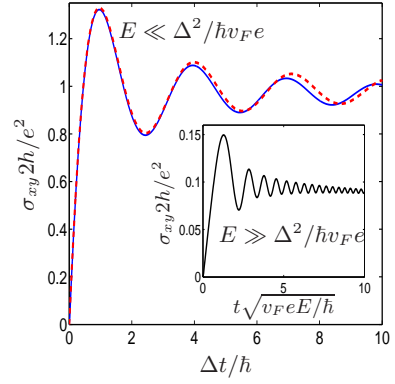}}
\caption{(Color online) The short time Hall conductivity is shown from Eq. \eqref{hallpert} (red dashed line) together with the numerical solution of Eq. \eqref{diractrans} (blue solid line) for weak electric field. The inset shows the numerical results for strong fields with the characteristic
oscillations set by the field. The transient response in both cases is well described by Eq. \eqref{transient}.
\label{shorttime}}
\end{figure}

We start (Fig. \ref{shorttime}) by analyzing its behaviour at weak electric fields, determined by the first term in Eq. 
\eqref{expr1}. 
The time dependent Hall current is
\begin{gather}
j_y(t)=\frac{e^2}{2h}\left[\frac{\Delta t}{\hbar}\left(\pi-2\textmd{Si}\left(\frac{2\Delta t}{\hbar}\right)\right) +2\sin^2\left(\frac{\Delta t}{\hbar}\right)\right]E,
\label{hallpert}
\end{gather}
where Si$(x)$ is the sine integral, and exhibits damped oscillations around the quantized value of the Hall conductivity as
$\sigma_{xy}=e^2/2h[1+\hbar\sin(2\Delta t/\hbar)/2\Delta t]$ with a frequency of $2\Delta/\hbar$.
Still at short times, but in the opposite small $\Delta$ and strong $E$ limit, similar oscillations with a frequency $\sim\sqrt{v_FeE/\hbar}$ show up in the response around the non-quantized asymptotic value. The transient behaviour at very short times rises linearly with $\Delta Et$ as
\begin{equation}
j_y(t)=\frac{e^2}{2h}\frac{\pi\Delta t}{\hbar}\textmd{min}(1,Wt/\hbar) E
\label{transient}
\end{equation}
with $W$ the cut-off, and is universal (no high energy modes involved) unless $t<\hbar/W$.

In the long time limit ($t\gg$min$[\hbar/\Delta,\sqrt{\hbar/v_FeE}]$), we can use the analogy of 
Eq. \eqref{diractrans} to the Landau-Zener problem\cite{vitanov,doraschwinger} of two-level crossing to 
determine $n_p(t)$: 
\begin{equation}
n_p(t)=\Theta(p_x(eEt-p_x))\exp\left(-\frac{\pi [(v_Fp_y)^2+\Delta^2]}{v_F\hbar eE}\right),
\label{pairprod}
\end{equation}
which is the pair production rate by Schwinger\cite{schwinger} and also the Landau Zener transition 
probability\cite{vitanov} between the initial and 
final levels, applicable if $(p_x,eEt-p_x) \gg \sqrt{p_y^2+(\Delta/v_F)^2}$. 

In the limit of long times, the second term in Eq. \eqref{expr1} dominates, and  the transverse current reaches a time independent value 
$j_y(t)=\sigma_{xy}E$
with
\begin{gather}
\sigma_{xy}=\frac{(ev_F)^2\Delta}{4\pi h}\int {d}{\bf p} \frac{1-2n_p(t)}{\varepsilon_p^3(t)}\approx\frac{e^2}{2h}\textmd{erf}
\left(\sqrt{\frac{\pi\Delta^2}{v_F\hbar eE}}\right),
\label{mainresult}
\end{gather}
which is our main result, erf$(x)$ being the error function.

Interestingly, the structure of the non-equilibrium Hall conductivity at long times agrees with the conventional equilibrium 
Kubo expression\cite{sinova,koshino} after shifting the 
momentum with the vector potential and replacing the equilibrium Fermi functions with the non-equilibrium momentum distribution, 
Eq. \eqref{pairprod}.
Alternatively, it reflects  the competition between Berry's curvature ($\Omega_p=v_F^2\Delta/2\varepsilon_p^3(t)$), protecting 
quantization and the difference of momentum distributions in the upper ($n_p(t)$) and lower 
($1-n_p(t)$) Dirac 
cones, spoiling it: when the two distributions are comparable due to tunneling, the gap becomes irrelevant, and the conductivity 
decays. In the limit of small fields ($E\ll \pi\Delta^2/v_F \hbar e$), we recover the quantized value 
\begin{equation}
\sigma_{xy}=\frac{e^2}{h}\int d{\bf p}\frac{\Omega_p}{2\pi}=\frac{e^2}{2h},
\end{equation}
without higher order perturbative or power law (in $E$) corrections. The additional terms contain the non-perturbative, exponential  
factor $\exp(-\pi\Delta^2/v_F\hbar eE)$, signaling the robustness
of Hall quantization \cite{avron} and the half-integer quantized magnetoelectric polarizability\cite{tme}.
In the strong field limit ($E\gg \pi\Delta^2/v_F \hbar e$),
it decays as
\begin{equation}
\sigma_{xy}=\frac{e^2}{2h}\frac{2\Delta}{\sqrt{v_F\hbar eE}}.
\label{hallstrong}
\end{equation}
For TI with a mass gap ($\Delta\neq 0$), the magnetization produced by surface currents probes the Hall conductivity through the topological magnetoelectric effect, and the
 magnetization \textit{parallel} to the electric field follows Eq. \eqref{mainresult}: its quantization breaks down with increasing field similarly to the Hall response.
When $\Delta=0$, the magnetization \textit{perpendicular} to $E$ becomes finite $\sim (e^2\pi/2h)E$ in weak fields\cite{doraschwinger}.

\begin{figure}[h!]
{\includegraphics[width=5cm,height=5cm]{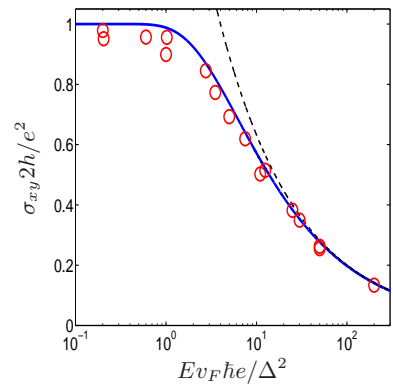}}

\caption{(Color online) The long time limit of the Hall conductivity is plotted as a function of the applied longitudinal electric field.
Quantization breaks down when $E\sim \Delta^2/\hbar v_F e$.
The red circles denote the numerical data from brute force integration
of the Dirac equation, Eq. \eqref{diractrans}, while the black dashed line is the approximate expression from Eq. \eqref{hallstrong} at 
large fields.
\label{nlfig}}
\end{figure}

Assuming a small gap of the order of 0.01-1~K (typical for the intrinsic SOC of graphene\cite{kanemele1,HuertasPRB2006} or TI) 
the crossover field is 0.001-10~V/m for $v_F\sim 10^6$~m/s, easily accessible experimentally. 
The Hall conductivity together with numerical results on the Dirac equation is shown in Fig. \ref{nlfig}.
The agreement between the analytically and numerically obtained conductivities is excellent.

We can get acquainted with the above results in different ways: first, a similar
situation occurs within equilibrium linear response (small $E$):  
the (spin-) Hall conductivity of massive Dirac electrons (i.e. graphene with intrinsic SOC and TI\cite{tihall}) is quantized to 
$e^2/2h$, 
when the chemical potential lies within the gap ($|\mu|<\Delta$). 
For $|\mu|>\Delta$, the chemical potential cuts into the continuum of band states,
and the conductivity decays as 
$e^2\Delta/2h|\mu|$, surviving even the effect of disorder\cite{sheng,sinitsyn}. 
Within our time-dependent formalism, the electric field can be thought formally as introducing an effective chemical potential 
$\mu_{eff}\sim\sqrt{\hbar v_FeE}$ (only $|v_Fp_y|<\mu_{eff}$ contributes), which upon substituting into the above
linear response expressions, parallels our findings.

Second, in the edge state picture, the Hall conductivity is provided by gapless, 
one-dimensional ballistic edge states, giving rise to the quantized value, which 
holds for weak electric fields. For strong fields, another type of gapless excitation starts
to contribute, due to tunneling between valence and conduction bands (Schwinger's 
pair production or Zener's dielectric breakdown), spoiling the perfect Hall quantization, as demonstrated above.

\begin{figure}[h!]
{\includegraphics[width=8cm,height=4.5cm]{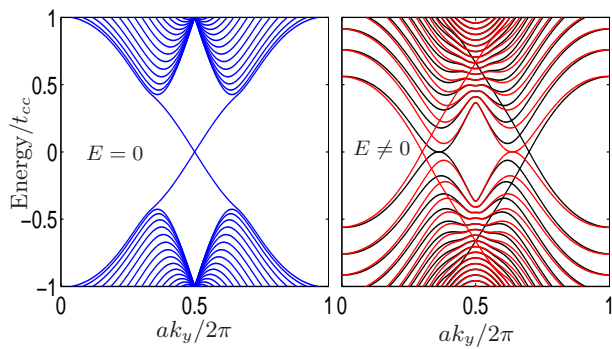}}
\caption{(Color online) The energy spectrum of a spin-Hall insulator\cite{kanemele1} in graphene, $t_{cc}$ is the hopping.
Left panel: without electric field, showing two gapless, spin degenerate edge states. Right panel: 
with finite critical electric field (red/black denoting up/down spin states), distorting the spectrum, 
and bringing additional levels into play around zero energy. Consequently, the spin-Hall conductivity is 
not quantized any more. Similar effects are generated by a strain induced pseudo-electric field having 
opposite sign in the two valleys, resulting in a valley-Hall effect. For stronger $E$, band crossing is more significant.
\label{halledge}}
\end{figure}

Another way to look at it is to consider the complementary stationary problem to Eq. \eqref{diraceq} of a static electric field 
in the form of a scalar potential ($\sim eEx$), and analyze the evolution of the
spectrum and edge states as a function of the electric field.
As a tight-binding example, we consider the spectrum of a zigzag graphene ribbon\cite{kanemele1} 
with intrinsic SOC, causing a gap with opposite sign between the two valleys, sublattices and spin directions in the continuum limit.
In the absence of an electric field, only the edge states, connecting the two Dirac cones, carry the transverse current, while
in a strong  electric field perpendicular to the edges, the effect of edge states is supplemented by the appearance of additional low 
energy modes living in two 
dimensions, due to the bands approaching each other, as seen in Fig.~\ref{halledge}.

At the same time, the longitudinal conductivity is  considerably enhanced in a 
strong electric field. In the time dependent framework, the increasing number
of electron-hole pairs, tunneled through the gap, facilitate longitudinal 
transport\cite{doraschwinger}, while in the stationary picture of Fig. \ref{halledge}, 
the presence of
new conducting channels around zero energy contribute the electric response. 
Note that the time-dependent framework provides a finite (spin-) Hall conductivity even 
in the absence of scattering\cite{sinova,sheng} due to its intrinsic character, not unlike the analysis of metallic 
graphene \cite{sinitsyn}, where additional disorder induced corrections were found, which we also expect to occur 
when scattering is added to our framework; these will also limit
the longitudinal conductivity\cite{doraschwinger}.

Third, 2D Dirac electrons in crossed stationary in plane electric ($E$) and perpendicular magnetic ($B$) field exhibit Landau 
quantization and 
subsequently quantized Hall conductivity. However, at $E=v_FB$, all Landau levels 
collapse\cite{llcollapse1,llcollapse2,llcollapse3} and should give way to a different 
Hall response. By defining the energy gap as the distance
between the Landau levels closest to the Dirac point, we get 
$\Delta_{Landau}=v_F\sqrt{2\hbar eB}$, yielding $E=\Delta_{Landau}^2/2 \hbar e v_F$ for the field causing the collapse of Landau levels, 
which agrees well with the crossover field where the (spin-) Hall response changes dramatically. 
We expect that some of our results can be transcribed to the quantum Hall breakdown in graphene\cite{singh}, 
testified by a Hall conductivity decreasing with the electric field, similarly to Eq. \eqref{hallstrong}.

These results are relatively robust against disorder,
because  the basic ingredient of the calculation is the non-equilibrium momentum 
distribution function, Eq. \eqref{pairprod},
which follows also from a semi-classical approach (WKB, expansion in $\hbar$).

We have studied the breakdown of the spin-Hall effect in graphene and surface Hall and magnetoelectric effect in topological insulators at nonzero electric field by concentrating 
on the real time dynamics of the transverse current.
The quantization of $\sigma_{xy}$, protected by the topologically invariant Chern number\cite{hasankane}, 
remains intact as long as the Hamiltonian varies smoothly (i.e. weak electric fields).
When non-adiabaticity enters via Landau-Zener transitions in strong field, quantization is lost and the Hall conductivity as well as the magnetoelectric coefficient decay as $E^{-1/2}$.


\begin{acknowledgments}
This work was supported by the Hungarian Scientific Research Fund No. K72613 and by the Bolyai program of the 
Hungarian Academy of Sciences.
\end{acknowledgments}

 
\bibliographystyle{apsrev}
\bibliography{refgraph}

\begin{thebibliography}{10}
\expandafter\ifx\csname bibnamefont\endcsname\relax
  \def\bibnamefont#1{#1}\fi
\expandafter\ifx\csname bibfnamefont\endcsname\relax
  \def\bibfnamefont#1{#1}\fi
\expandafter\ifx\csname url\endcsname\relax
  \def\url#1{\texttt{#1}}\fi
\expandafter\ifx\csname urlprefix\endcsname\relax\def\urlprefix{URL }\fi
\providecommand{\bibinfo}[2]{#2}
\providecommand{\eprint}[2][]{\url{#2}}

\bibitem{castro07}
\bibinfo{author}{\bibfnamefont{A.~H.} \bibnamefont{{Castro Neto}}},
  \bibinfo{author}{\bibfnamefont{F.}~\bibnamefont{Guinea}},
  \bibinfo{author}{\bibfnamefont{N.~M.~R.} \bibnamefont{Peres}},
  \bibinfo{author}{\bibfnamefont{K.~S.} \bibnamefont{Novoselov}},
  \bibnamefont{and} \bibinfo{author}{\bibfnamefont{A.~K.} \bibnamefont{Geim}},
  \bibinfo{journal}{Rev. Mod. Phys.} \textbf{\bibinfo{volume}{81}},
  \bibinfo{pages}{109} (\bibinfo{year}{2009}).

\bibitem{roomhallgraphene}
\bibinfo{author}{\bibfnamefont{K.~S.} \bibnamefont{Novoselov}},
  \bibinfo{author}{\bibfnamefont{Z.}~\bibnamefont{Jiang}},
  \bibinfo{author}{\bibfnamefont{Y.}~\bibnamefont{Zhang}},
  \bibinfo{author}{\bibfnamefont{S.~V.} \bibnamefont{Morozov}},
  \bibinfo{author}{\bibfnamefont{H.~L.} \bibnamefont{Stormer}},
  \bibinfo{author}{\bibfnamefont{U.}~\bibnamefont{Zeitler}},
  \bibinfo{author}{\bibfnamefont{J.~C.} \bibnamefont{Maan}},
  \bibinfo{author}{\bibfnamefont{G.~S.} \bibnamefont{Boebinger}},
  \bibinfo{author}{\bibfnamefont{P.}~\bibnamefont{Kim}}, \bibnamefont{and}
  \bibinfo{author}{\bibfnamefont{A.~K.} \bibnamefont{Geim}},
  \bibinfo{journal}{Science} \textbf{\bibinfo{volume}{315}},
  \bibinfo{pages}{1379} (\bibinfo{year}{2007}).

\bibitem{kanemele1}
\bibinfo{author}{\bibfnamefont{C.~L.} \bibnamefont{Kane}} \bibnamefont{and}
  \bibinfo{author}{\bibfnamefont{E.~J.} \bibnamefont{Mele}},
  \bibinfo{journal}{Phys. Rev. Lett.} \textbf{\bibinfo{volume}{95}},
  \bibinfo{pages}{226801} (\bibinfo{year}{2005}).

\bibitem{hasankane}
\bibinfo{author}{\bibfnamefont{M.~Z.} \bibnamefont{Hasan}} \bibnamefont{and}
  \bibinfo{author}{\bibfnamefont{C.~L.} \bibnamefont{Kane}},
  \bibinfo{note}{{a}rXiv:1002.3895}.

\bibitem{bernevig}
\bibinfo{author}{\bibfnamefont{B.~A.} \bibnamefont{Bernevig}},
  \bibinfo{author}{\bibfnamefont{T.~L.} \bibnamefont{Hughes}},
  \bibnamefont{and} \bibinfo{author}{\bibfnamefont{S.-C.} \bibnamefont{Zhang}},
  \bibinfo{journal}{Science} \textbf{\bibinfo{volume}{314}},
  \bibinfo{pages}{1757} (\bibinfo{year}{2006}).

\bibitem{phystoday}
\bibinfo{author}{\bibfnamefont{X.-L.} \bibnamefont{Qi}} \bibnamefont{and}
  \bibinfo{author}{\bibfnamefont{S.-C.} \bibnamefont{Zhang}},
  \bibinfo{journal}{Phys. Today} \textbf{\bibinfo{volume}{63}},
  \bibinfo{pages}{33} (\bibinfo{year}{2010}).

\bibitem{tme}
\bibinfo{author}{\bibfnamefont{X.-L.} \bibnamefont{Qi}},
  \bibinfo{author}{\bibfnamefont{T.~L.} \bibnamefont{Hughes}},
  \bibnamefont{and} \bibinfo{author}{\bibfnamefont{S.-C.} \bibnamefont{Zhang}},
  \bibinfo{journal}{Phys. Rev. B} \textbf{\bibinfo{volume}{78}},
  \bibinfo{pages}{195424} (\bibinfo{year}{2008}).

\bibitem{ludwig94}
\bibinfo{author}{\bibfnamefont{A.~W.~W.} \bibnamefont{Ludwig}},
  \bibinfo{author}{\bibfnamefont{M.~P.~A.} \bibnamefont{Fisher}},
  \bibinfo{author}{\bibfnamefont{R.}~\bibnamefont{Shankar}}, \bibnamefont{and}
  \bibinfo{author}{\bibfnamefont{G.}~\bibnamefont{Grinstein}},
  \bibinfo{journal}{Phys. Rev. B} \textbf{\bibinfo{volume}{50}},
  \bibinfo{pages}{7526} (\bibinfo{year}{1994}).

\bibitem{doraschwinger}
\bibinfo{author}{\bibfnamefont{B.}~\bibnamefont{D\'ora}} \bibnamefont{and}
  \bibinfo{author}{\bibfnamefont{R.}~\bibnamefont{Moessner}},
  \bibinfo{journal}{Phys. Rev. B} \textbf{\bibinfo{volume}{81}},
  \bibinfo{pages}{165431} (\bibinfo{year}{2010}).

\bibitem{beenakker}
\bibinfo{author}{\bibfnamefont{C.~W.~J.} \bibnamefont{Beenakker}},
  \bibinfo{journal}{Rev. Mod. Phys.} \textbf{\bibinfo{volume}{80}},
  \bibinfo{pages}{1337} (\bibinfo{year}{2008}).

\bibitem{schwinger}
\bibinfo{author}{\bibfnamefont{J.}~\bibnamefont{Schwinger}},
  \bibinfo{journal}{Phys. Rev.} \textbf{\bibinfo{volume}{82}},
  \bibinfo{pages}{664} (\bibinfo{year}{1951}).

\bibitem{allor}
\bibinfo{author}{\bibfnamefont{D.}~\bibnamefont{Allor}},
  \bibinfo{author}{\bibfnamefont{T.~D.} \bibnamefont{Cohen}}, \bibnamefont{and}
  \bibinfo{author}{\bibfnamefont{D.~A.} \bibnamefont{McGady}},
  \bibinfo{journal}{Phys. Rev. D} \textbf{\bibinfo{volume}{78}},
  \bibinfo{pages}{096009} (\bibinfo{year}{2008}).

\bibitem{rosenstein}
\bibinfo{author}{\bibfnamefont{R.}~\bibnamefont{Rosenstein}},
  \bibinfo{author}{\bibfnamefont{M.}~\bibnamefont{Lewkowicz}},
  \bibinfo{author}{\bibfnamefont{H.~C.} \bibnamefont{Kao}}, \bibnamefont{and}
  \bibinfo{author}{\bibfnamefont{Y.}~\bibnamefont{Korniyenko}},
  \bibinfo{journal}{Phys. Rev. B} \textbf{\bibinfo{volume}{81}},
  \bibinfo{pages}{041416} (\bibinfo{year}{2010}).

\bibitem{mauri}
\bibinfo{author}{\bibfnamefont{N.}~\bibnamefont{Vandecasteele}},
  \bibinfo{author}{\bibfnamefont{A.}~\bibnamefont{Barreiro}},
  \bibinfo{author}{\bibfnamefont{M.}~\bibnamefont{Lazzeri}},
  \bibinfo{author}{\bibfnamefont{A.}~\bibnamefont{Bachtold}}, \bibnamefont{and}
  \bibinfo{author}{\bibfnamefont{F.}~\bibnamefont{Mauri}},
  \bibinfo{journal}{Phys. Rev. B} \textbf{\bibinfo{volume}{82}},
  \bibinfo{pages}{045416} (\bibinfo{year}{2010}).

\bibitem{singh}
\bibinfo{author}{\bibfnamefont{V.}~\bibnamefont{Singh}} \bibnamefont{and}
  \bibinfo{author}{\bibfnamefont{M.~M.} \bibnamefont{Deshmukh}},
  \bibinfo{journal}{Phys. Rev. B} \textbf{\bibinfo{volume}{80}},
  \bibinfo{pages}{081404} (\bibinfo{year}{2009}).

\bibitem{franz}
\bibinfo{author}{\bibfnamefont{I.}~\bibnamefont{Garate}} \bibnamefont{and}
  \bibinfo{author}{\bibfnamefont{M.}~\bibnamefont{Franz}},
  \bibinfo{journal}{Phys. Rev. Lett.} \textbf{\bibinfo{volume}{104}},
  \bibinfo{pages}{146802} (\bibinfo{year}{2010}).

\bibitem{vitanov}
\bibinfo{author}{\bibfnamefont{N.~V.} \bibnamefont{Vitanov}} \bibnamefont{and}
  \bibinfo{author}{\bibfnamefont{B.~M.} \bibnamefont{Garraway}},
  \bibinfo{journal}{Phys. Rev. A} \textbf{\bibinfo{volume}{53}},
  \bibinfo{pages}{4288} (\bibinfo{year}{1996}).

\bibitem{koshino}
\bibinfo{author}{\bibfnamefont{M.}~\bibnamefont{Koshino}},
  \bibinfo{journal}{Phys. Rev. B} \textbf{\bibinfo{volume}{78}},
  \bibinfo{pages}{155411} (\bibinfo{year}{2008}).

\bibitem{sinova}
\bibinfo{author}{\bibfnamefont{J.}~\bibnamefont{Sinova}},
  \bibinfo{author}{\bibfnamefont{D.}~\bibnamefont{Culcer}},
  \bibinfo{author}{\bibfnamefont{Q.}~\bibnamefont{Niu}},
  \bibinfo{author}{\bibfnamefont{N.~A.} \bibnamefont{Sinitsyn}},
  \bibinfo{author}{\bibfnamefont{T.}~\bibnamefont{Jungwirth}},
  \bibnamefont{and} \bibinfo{author}{\bibfnamefont{A.~H.}
  \bibnamefont{MacDonald}}, \bibinfo{journal}{Phys. Rev. Lett.}
  \textbf{\bibinfo{volume}{92}}, \bibinfo{pages}{126603}
  (\bibinfo{year}{2004}).

\bibitem{avron}
\bibinfo{author}{\bibfnamefont{J.~E.} \bibnamefont{Avron}} \bibnamefont{and}
  \bibinfo{author}{\bibfnamefont{Z.}~\bibnamefont{Kons}}, \bibinfo{journal}{J.
  Phys. A: Math. Gen.} \textbf{\bibinfo{volume}{32}}, \bibinfo{pages}{6097}
  (\bibinfo{year}{1999}).

\bibitem{HuertasPRB2006}
\bibinfo{author}{\bibfnamefont{D.}~\bibnamefont{Huertas-Hernando}},
  \bibinfo{author}{\bibfnamefont{F.}~\bibnamefont{Guinea}}, \bibnamefont{and}
  \bibinfo{author}{\bibfnamefont{A.}~\bibnamefont{Brataas}},
  \bibinfo{journal}{Phys. Rev. B} \textbf{\bibinfo{volume}{74}},
  \bibinfo{pages}{155426} (\bibinfo{year}{2006}).

\bibitem{tihall}
\bibinfo{author}{\bibfnamefont{J.}~\bibnamefont{Zang}} \bibnamefont{and}
  \bibinfo{author}{\bibfnamefont{N.}~\bibnamefont{Nagaosa}},
  \bibinfo{journal}{Phys. Rev. B} \textbf{\bibinfo{volume}{81}},
  \bibinfo{pages}{245125} (\bibinfo{year}{2010}).

\bibitem{sheng}
\bibinfo{author}{\bibfnamefont{L.}~\bibnamefont{Sheng}},
  \bibinfo{author}{\bibfnamefont{D.~N.} \bibnamefont{Sheng}},
  \bibinfo{author}{\bibfnamefont{C.~S.} \bibnamefont{Ting}}, \bibnamefont{and}
  \bibinfo{author}{\bibfnamefont{F.~D.~M.} \bibnamefont{Haldane}},
  \bibinfo{journal}{Phys. Rev. Lett.} \textbf{\bibinfo{volume}{95}},
  \bibinfo{pages}{136602} (\bibinfo{year}{2005}).

\bibitem{sinitsyn}
\bibinfo{author}{\bibfnamefont{N.~A.} \bibnamefont{Sinitsyn}},
  \bibinfo{author}{\bibfnamefont{J.~E.} \bibnamefont{Hill}},
  \bibinfo{author}{\bibfnamefont{H.}~\bibnamefont{Min}},
  \bibinfo{author}{\bibfnamefont{J.}~\bibnamefont{Sinova}}, \bibnamefont{and}
  \bibinfo{author}{\bibfnamefont{A.~H.} \bibnamefont{MacDonald}},
  \bibinfo{journal}{Phys. Rev. Lett.} \textbf{\bibinfo{volume}{97}},
  \bibinfo{pages}{106804} (\bibinfo{year}{2006}).

\bibitem{llcollapse1}
\bibinfo{author}{\bibfnamefont{V.}~\bibnamefont{Lukose}},
  \bibinfo{author}{\bibfnamefont{R.}~\bibnamefont{Shankar}}, \bibnamefont{and}
  \bibinfo{author}{\bibfnamefont{G.}~\bibnamefont{Baskaran}},
  \bibinfo{journal}{Phys. Rev. Lett.} \textbf{\bibinfo{volume}{98}},
  \bibinfo{pages}{116802} (\bibinfo{year}{2007}).

\bibitem{llcollapse2}
\bibinfo{author}{\bibfnamefont{N.~M.~R.} \bibnamefont{Peres}} \bibnamefont{and}
  \bibinfo{author}{\bibfnamefont{E.~V.} \bibnamefont{Castro}},
  \bibinfo{journal}{J. Phys. Cond. Matter} \textbf{\bibinfo{volume}{19}},
  \bibinfo{pages}{406231} (\bibinfo{year}{2007}).

\bibitem{llcollapse3}
\bibinfo{author}{\bibfnamefont{S.}~\bibnamefont{Mondal}},
  \bibinfo{author}{\bibfnamefont{D.}~\bibnamefont{Sen}},
  \bibinfo{author}{\bibfnamefont{K.}~\bibnamefont{Sengupta}}, \bibnamefont{and}
  \bibinfo{author}{\bibfnamefont{R.}~\bibnamefont{Shankar}},
  \bibinfo{journal}{Phys. Rev. B} \textbf{\bibinfo{volume}{82}},
  \bibinfo{pages}{045120} (\bibinfo{year}{2010}).

\end{thebibliography}
\end{document}